\documentclass[%
 reprint,
 amsmath,amssymb,
 aps,
 prl,
]{revtex4-2}

\usepackage{aas_macros}
\usepackage{graphicx}
\usepackage{dcolumn}
\usepackage{bm}
\usepackage{hyperref}
\usepackage{color}
\usepackage{natbib}
\usepackage{CJK}
\usepackage[normalem]{ulem}
\usepackage[english]{babel}

\renewcommand{\prl}{Phys.~Rev.~Lett.}

\newcommand{\cqg}{{Class.~Quant.~Grav.}~}
\renewcommand{\apj}{{Astrophys.~J.}}
\renewcommand{\apjl}{{Astrophys.~J.~Lett.~}}
\renewcommand{\apjs}{{Astrophys.~J.~Suppl.~Ser.~}}

\renewcommand{\nat}{Nature}
\newcommand{\natrevphys}{Nat.~Rev.~Phys.~}
\newcommand{\natastron}{Nat.~Astron.~}
\renewcommand{\mnras}{{Mon.~Not.~R.~Astron.~Soc.~}}
\renewcommand{\aap}{Astron.~Astrophys.~}

\begin{document}
\begin{CJK*}{UTF8}{gbsn} 
\preprint{APS/123-QED}


\title{Jet-driven explosion of an accretion-induced white-dwarf collapse via a magnetorotational dynamo}

\author{Luciano Combi$^{1,2}$}

\author{Daniel M.~Siegel$^{3,1,2}$}

\author{Brian D.~Metzger$^{4,5}$}

\affiliation{$^1$Perimeter Institute for Theoretical Physics, Waterloo, Ontario N2L 2Y5, Canada\\
$^2$Department of Physics, University of Guelph, Guelph, Ontario N1G 2W1, Canada\\
$^3$Institute of Physics, University of Greifswald, D-17489 Greifswald, Germany\\
$^4$Department of Physics and Columbia Astrophysics Laboratory, Columbia University, New York, NY 10027, USA\\
$^5$Center for Computational Astrophysics, Flatiron Institute, 162 5th Ave, New York, NY 10010, USA}



\begin{abstract}
 The accretion-induced collapse (AIC) of a rotating white dwarf (WD) offers a potential site of millisecond pulsars/magnetars, gamma-ray bursts, and r-process nucleosynthesis. We present three-dimensional general-relativistic magneto-hydrodynamical simulations including neutrinos of magnetorotational AIC, assuming the WD is rapidly spinning with a weak magnetic field confined below its surface (likely a prerequisite for rapid rotation). Within milliseconds after core bounce, the magnetic field is exponentially amplified near the surface of the proto-neutron star (PNS). We witness the emergence of a small-scale turbulent and mean-field, large-scale MRI-driven dynamo in the neutrino-cooled centrifugally supported disk formed around the PNS, which generates bundles of large-scale toroidal field with alternating polarity. The amplified field becomes buoyant and is advected above the PNS, generating a magnetic tower that drives a mildly relativistic striped jet. The jet breaks out of the WD, clearing the way for a powerful magnetized neutron-rich wind from the disk. Although our simulation cannot follow the long-term Kelvin-Helmholtz cooling phase of the PNS, the conditions are ripe for the formation of a GRB powered by magnetar spin-down. A similar dynamo may operate in magnetorotational core-collapse supernovae and neutron-star mergers.
\end{abstract}

\maketitle
\end{CJK*}


\textit{Introduction.---}The enormous binding energy suddenly released in a stellar collapse ($E\sim\!10^{53}$\,erg) can power the most energetic transients in the Universe. Large-scale magnetic fields can tap the available gravitational and rotational energy to produce $\gamma$-ray bursts (GRBs; \cite{woosley1993GammaRay,blackman_fueling_1998,woosley_central_1999, gottlieb_shes_2024,thompson1994Model,metzger_protomagnetar_2011}), power hyperenergetic supernovae Ic-bl \cite{leblanc_numerical_1970,mosta2014Magnetorotational, kuroda_magnetorotational_2020, dessart_proto-neutron_2008, obergaulinger_magnetorotational_2020} or superluminous supernovae \cite{kasen_supernova_2010,woosley_bright_2010}, and launch magnetohydrodynamic (MHD) winds from the remnant that give rise to kilonovae \cite{metzger2007ProtoNeutron, metzger2018Magnetar, desai_three-dimensional_2023}. Such fields can extract neutron-rich material from the resulting nascent proto-neutron star (PNS) or black hole fast enough to trigger rapid neutron capture (the r-process) in the outflowing plasma \cite{halevi_r-process_2018,reichert_nucleosynthesis_2021,issa_magnetically_2025}, which makes magnetorotational stellar collapse, in addition to neutron-star mergers, a leading candidate for the synthesis of heavy elements in the universe \cite{cowan_origin_2021,siegel_r-process_2022}.

The amplification of magnetic fields in the progenitor through non-linear dynamo processes during and after collapse is key to understanding the explosion mechanism itself and predicting electromagnetic radiation and nucleosynthesis yields from these violent events, as well as the origin of magnetars.  Mechanisms to generate strong magnetic fields in PNS range from compression of flux-frozen relic fields \citep{schneider_stellar_2019}, Tayler instability \citep{Skoutnev:2024vlm, Reboul-Salze:2024jst}, convection \citep{thompson1993neutron, raynaud_magnetar_2020}, to magnetorotational instability (MRI)-driven dynamos \citep{obergaulinger_semi-global_2009, guilet2015numerical, reboul2025mri, reboul-salze_mri-driven_2022, mosta_large-scale_2015, akiyama2003magnetorotational}. The latter are particularly effective in rotating PNS and may generate sufficiently strong fields to power a successful magnetorotational explosion in some hypernovae \citep{obergaulinger_axisymmetric_2006, mosta_magnetorotational_2014, bugli_impact_2020}.

White dwarfs (WD) with ONeMg cores can undergo accretion-induced collapse (AIC) to a neutron star when approaching the Chandrasekhar limit and thus represent alternative progenitors of PNS and magnetars as well as potential sites for r-process nucleosynthesis \cite{wheeler_r-process_1998,fryer_what_1999}. A WD may accrete sufficient angular momentum and mass from a companion star via Roche-lobe overflow \cite{Tauris+13,Brooks+17} or through merger with another WD in a double WD binary \cite{Yoon+07} to collapse to a rapidly rotating PNS, surrounded by a centrifugally supported accretion disk \citep{dessart_magnetically_2007, dessart_multidimensional_2006, abdikamalov_axisymmetric_2009, Micchi+23}. Winds launched from the disk may release neutron-rich material conducive to r-process nucleosynthesis with associated kilonova-type emission (similar to the aftermath of neutron-star mergers; \cite{siegel_three-dimensional_2017}) and $^{56}$Ni-rich outflows \citep{metzger2009nickel}. Two-dimensional simulations, which are subject to the anti-dynamo theorem in axisymmetry \cite{cowling_magnetic_1933}, suggest that, assuming a sufficiently strong, large-scale magnetic field anchored to the vicinity of the nascent PNS, AICs are also potential sources for GRBs \citep{dessart_magnetically_2007, cheong_gamma-ray_2025}.

Despite these appealing features as sources of r-process elements, GRBs, and magnetars, `magnetorotational' AIC models face challenges. The necessary magnetic field strengths to form a magnetar through flux conservation are at least a factor $\sim\!100$ larger than the most strongly magnetized observed WDs ($B\lesssim 10^9$\,G \cite{ferrario_magnetic_2005}). Furthermore, rapid rotation and a large-scale magnetic field are mutually exclusive for an accreting star, since magnetic braking in the magnetosphere limits the accreted angular momentum \citep{thompson95}. For a rapidly rotating WD to form, the seed magnetic field should be buried within the star. Although this might be expected from a dynamo occurring during the crystallization of the WD core \citep{schreiber_origin_2021,ginzburg2022slow}, the flux-compressed field ($\sim\!10^8$\,G) is not sufficiently strong to instantly emerge from the PNS surface and have a dynamical impact on the explosion.



Here, we present evidence of an MRI-driven turbulent and mean-field $\alpha\Omega$ dynamo operating in the accretion disk surrounding the PNS, which provides sufficient large-scale flux to generate a magnetic tower anchored above a convective envelope of the PNS, capable of a successful jet-driven magnetorotational explosion. We demonstrate that early amplification by the MRI around the PNS core, similar to the magnetorotational core-collapse of massive stars \cite{mosta_large-scale_2015}, leads to rapid redistribution of angular momentum, and is likely incapable of launching a successful jet. The jet-driven explosion in the polar regions clears the path to neutron-rich MHD winds from the disk. If the large-scale flux remains attached to the star as the disk mass is accreted and the PNS cools down, we argue that the jet can reach high enough magnetization on $\sim$\,seconds to power a GRB.



\textit{Computational setup.---}We solve the ideal general-relativistic MHD equations in full 3D with neutrino interactions, coupled to Einstein's field equations, using a modified version \cite{siegel_three-dimensional_2018,combi_grmhd_2023} of the open-source, flux-conservative \texttt{GRHydro} code \cite{mosta_grhydro_2014}, which is part of the Einstein Toolkit \cite{loffler_einstein_2012}. We use the Carpet driver \cite{schnetter_evolutions_2004} for Berger-Oliger mesh refinement with subcycling in time. The Cartesian grid hierarchy consists of ten levels box-in-box of extent $(8000\,{\rm km})^3$, with a finest resolution of $\Delta x=240\,{\rm m}$ in the 20\,km box.  

The initial data constructed with \texttt{RNS} \cite{stergioulas_comparing_1995} consists of a WD near its Chandrasekhar mass, $M \simeq 1.45\,{M_{\odot}}$, rigidly rotating near its mass-shedding limit, with an equatorial (polar) radius of $1700$\,km ($1122$\,km). We assume that the star is cold ($T\approx 0.01\,{\rm MeV}$) and that electron captures have not set in yet (proton fraction $Y_e=0.5$; these assumptions may not hold for a WD-WD merger \cite{schwab_evolution_2016}). 
We place a small poloidal magnetic loop in the star's core using a vector potential with the only non-vanishing component $A_{\phi}=A_{b}{\rm min}(p-0.04p_{\rm max},0)$. Here, $A_b$ sets the maximum strength of the field to $B=10^{12}$\,G and $p_{\rm max}$ is the maximum pressure in the star. The initial magnetic energy ($\approx\!10^{46}\,{\rm erg}$) is mostly contained within a radius of $200\,{\rm km}$, which is contracted to \emph{within} the nascent PNS upon collapse. 
Although the extrapolated surface value is $\ll10^{6}$\,G, we need to assume a relatively strong seed field near the WD core to resolve the MRI shortly after PNS formation, given the finite resolution set by computational constraints. As we argue (see End Matter), much smaller seed field strengths nevertheless produce similar results. 

We model electron captures onto nuclei, which are important near the WD core $(\rho \sim 10^{10}\,{\rm g}\,{\rm cm}^{-3})$ during the collapse, by the prescription $Y_{\rm e}(\rho)$ of Ref.~\citep{liebendorfer_simple_2005}. 
We switch on weak interactions at bounce with an approximate, one-moment (M0) neutrino transport scheme adapted from Refs.~\citep{radice2016Dynamical,radice2018Binary}. We simulate a total duration of $250\,{\rm ms}$ post-bounce.


\begin{figure}
    \centering    
    \includegraphics[width=0.9\linewidth]{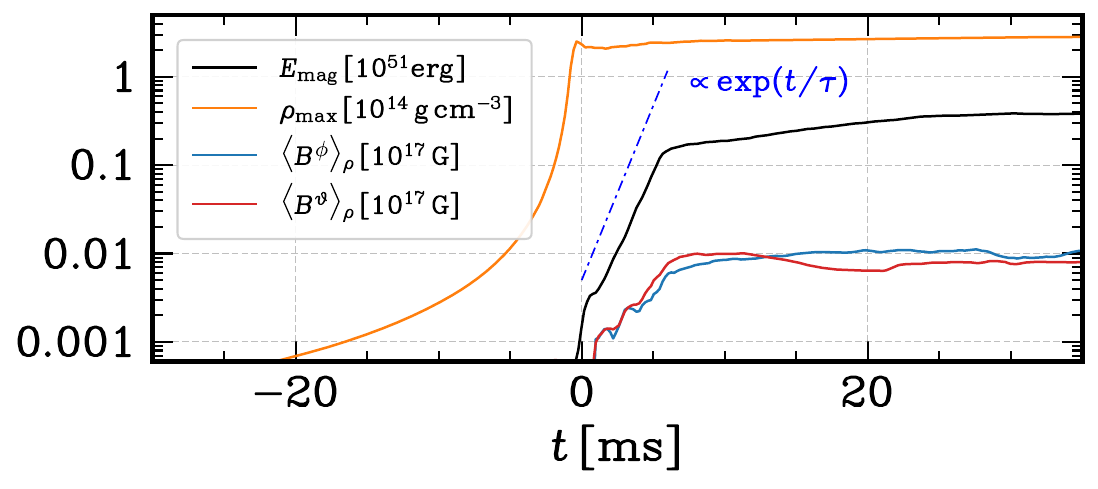}
    \includegraphics[width=0.9\linewidth]{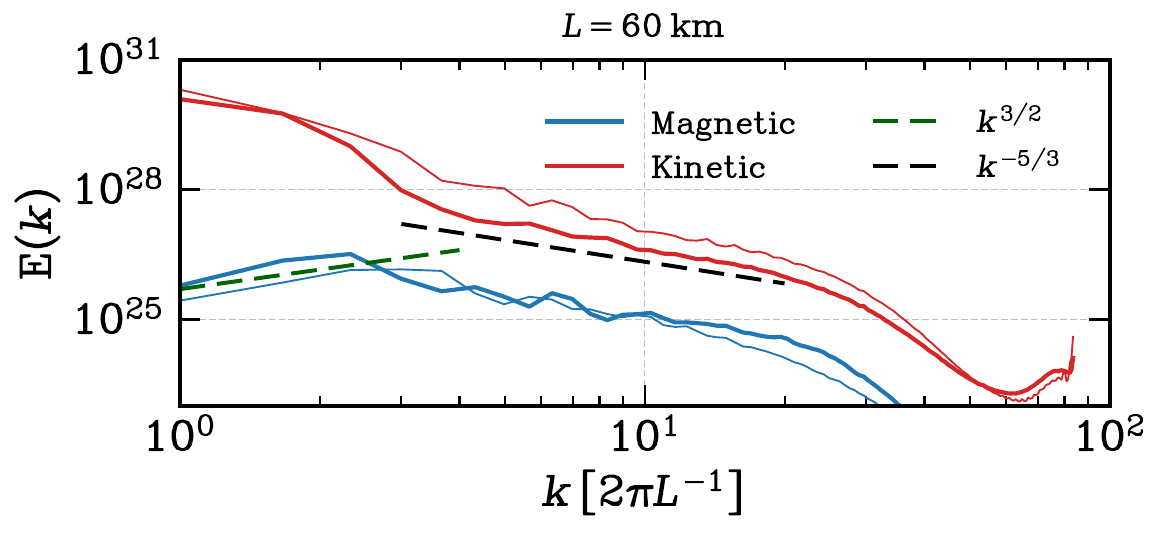}
    \caption{Top: evolution during collapse $(t<0)$ and post-bounce $(t>0)$ of the maximum rest-max density $\rho_{\rm max}$, the total magnetic energy $E_{\rm mag}$, and the spherically averaged, density-weighted toroidal $(\langle B^{\phi}\rangle_{\rho})$ and poloidal $(\langle B^{\theta}\rangle_{\rho})$ magnetic field strengths in the region $r\in (15,25)\,{\rm km}$. The dashed-dotted line indicates exponential growth with a timescale $\tau = 1/\Omega(20{\rm km})$. 
    Bottom: total magnetic and kinetic energy density spectrum after bounce at $t\approx 14\,{\rm ms}$ (thick lines) and at $t\approx 160\,{\rm ms}$ (thin lines). The spectra are normalized with an offset for better readability.}
    \label{fig:ek_evolmag}
\end{figure}

\begin{figure}
    \centering
    \includegraphics[width=1.0\linewidth]{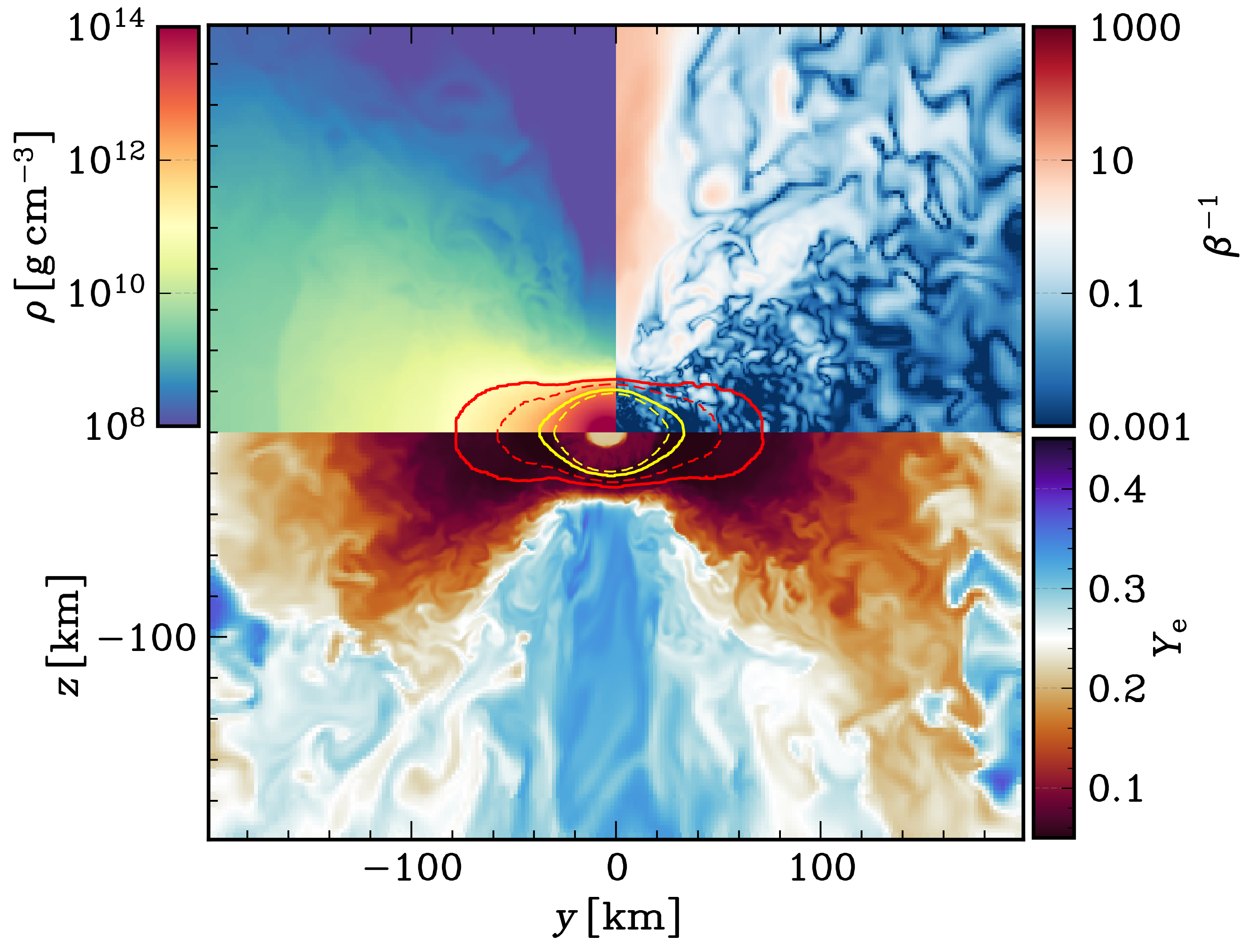}
    \includegraphics[width=1.0\linewidth]{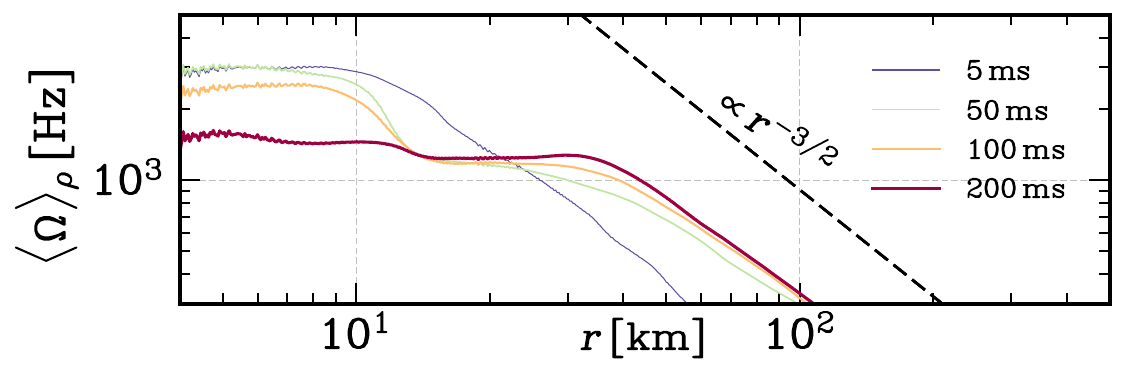}
    \caption{Top: 2D snapshot in the $xz$-plane at $t=150\,{\rm ms}$ showing rest-mass density ($\rho$, top left), inverse plasma beta ($\beta^{-1}$, top right), and proton fraction ($Y_{\rm e}$, bottom). 
    The red and yellow thick (dashed) contours represent optical depths to electron (anti-)neutrinos of $\tau = 1$ and $\tau = 20$, respectively. Bottom: spherically averaged, density-weighted angular velocity as a function of radius at different epochs post-bounce. 
    }
    \label{fig:xz}
\end{figure}

\textit{Magnetic field amplification---}Pressure support in the core of the rotating WD is decreased due to electron captures, and collapse ensues for $\approx\!66\,{\rm ms}$ until nuclear densities are reached ($\approx\! 3\times10^{14}\,{\rm g}\,{\rm cm}^{-3}$), the EOS stiffens, and the infalling gas bounces off the newly formed PNS (marked by a kink in the evolution of the maximum rest-mass density, Fig.~\ref{fig:ek_evolmag}). We denote the moment of bounce by $t=0\,{\rm ms}$. Conservation of angular momentum during collapse leads to a shear flow around the PNS after bounce, with a nearly rigidly rotating PNS core ($r\lesssim 10\,{\rm km}$) at initial angular velocity of $\Omega\approx 3\,{\rm kHz}$, transitioning into a rotationally supported surrounding flow, $\Omega \propto r^{-3/2}$ (Fig.~\ref{fig:xz}). The seed magnetic field is compressed and amplified during collapse due to flux conservation, reaching up to $B\sim 10^{14}\,{\rm G}$ ($B\sim 10^{15}\,{\rm G}$) in the shear layer around (core of) the PNS. The field remains dynamically irrelevant throughout the collapse with $\beta^{-1}=b^2/(2p)\lesssim10^{-3}$, where $b$ is the comoving magnetic field strength (see End Matter).

Upon bounce, the magnetic field is rapidly amplified by a turbulent dynamo. The total magnetic energy increases exponentially by two orders of magnitude on a timescale of $\tau\approx 1.1\,{\rm ms}\approx 1/\Omega(r=20\,{\rm km})$, plateauing within $t\approx 5\,{\rm ms}$ and reaching $E_{\rm mag}\approx 10^{49}\,{\rm erg}$ by $t\approx 20\,{\rm ms}$, with $\beta^{-1} \sim 0.1$ outside the PNS core (Fig.~\ref{fig:ek_evolmag}). As a result of this amplification phase, the fluid becomes turbulent, with the kinetic energy spectrum approaching a Kolmogorov power-law $\propto k^{-5/3}$ between $k\approx 3-30$ (Fig.~\ref{fig:ek_evolmag}), where $k$ is the wavenumber in units of $2\pi/L$ and $L=70\,{\rm km}$ the length of the turbulent box considered here. The magnetic energy increases roughly $\propto k^{3/2}$ as expected for the Kazantsev solution, consistent with kinetic Kolmogorov turbulence \cite{kazantsev_enhancement_1968}, reaching a peak at $k\approx 4$ close to equipartition with the kinetic energy, and approaching Kolmogorov's decay at larger $k$. This behavior indicates the presence of a small-scale dynamo with inverse MHD cascade in a regime in which the initial magnetic energy is smaller than the kinetic energy, and the magnetic Prandtl number is of order unity (i.e., resistivity and numerical viscosity are controlled by the grid scale). Given $d\Omega/dr < 0$, the initial weak seed field, and the agreement with the expected exponential amplification timescale $\propto \Omega^{-1}$ (see above), the dynamo outside the PNS core is likely driven by the MRI, similar to magnetorotational core-collapse supernovae \cite{mosta_large-scale_2015}. Prompt convection behind the bounce shock also stirs turbulence in the PNS and could help amplify the field at early times, even without rotation. As in Ref.~\cite{mosta_large-scale_2015}, we observe the formation of a large-scale toroidal field around the PNS core, which, however, does not promptly emerge from the PNS vicinity (except when a larger seed magnetic field strength of $B=10^{13}\,{\rm G}$ is used \cite{combi2025aiclong}). Because rapid redistribution of angular momentum within $\sim\!40$\,ms renders the PNS stable to the instability (see below), the accretion disk is a more promising site to amplify the magnetic field, explode the star, and sustain a Poynting flux from the PNS on timescales of interest. 

\begin{figure}
    \centering    \includegraphics[width=1.0\linewidth]{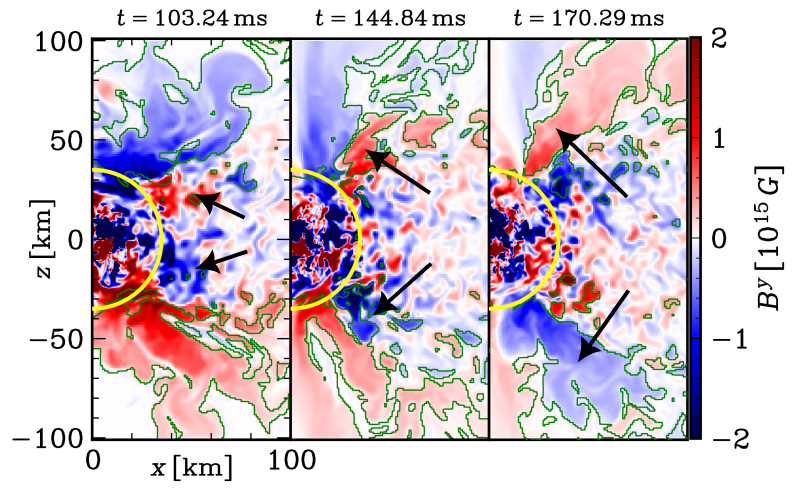}
    \includegraphics[width=1.0\linewidth]{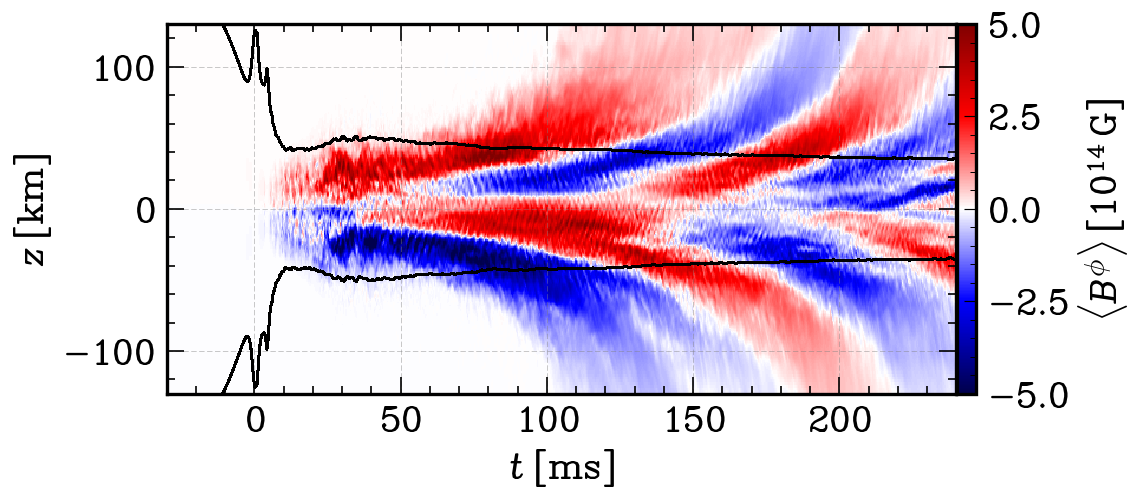}
    \caption{Top: snapshots of $B^{y}$ in the $xz$-plane (toroidal magnetic field). Field bundles of coherent polarity emerge quasi-periodically with alternating polarity from the accretion disk, accumulate above the PNS, and become buoyant when $\beta^{-1}\gtrsim1$ (black arrows track movement, green lines mark $\beta^{-1}\gtrsim0.1$). They partially reconnect in the polar cap, generating a `striped' jet. The yellow line marks the boundary of the convective PNS envelope at $r\approx 35\,{\rm km}$. 
    Bottom: butterfly (spacetime) diagram of the azimuthally averaged toroidal field, radially averaged between $50-60$\,km. Black lines mark twice the scale-height, $2H$.}
    \label{fig:dynamo}
\end{figure}

As the bounce shock propagates outward from the surface of the PNS core, it dissociates and deleptonizes the infalling material and leaves behind a neutron-rich, rotationally-supported ($\Omega\propto r^{-3/2}$) accretion disk extending out to $r\sim 150$\,km (Fig.~\ref{fig:xz}). 
The disk further neutronizes due to electron captures on protons, with the degenerate plasma maintaining a high neutron richness $Y_e \sim 0.1$ on timescales of interest as a result of a self-regulation mechanism due to Pauli blocking of $e^{+}$ captures on neutrons, similar to collapsar and neutron-star post-merger accretion disks \cite{siegel_three-dimensional_2017,siegel_collapsars_2019,agarwal_ignition_2025}. The disk smoothly joins a nearly hydrostatic envelope around the PNS at $r\approx35\,{\rm km}$ (Fig.~\ref{fig:xz}), which is optically thick to neutrinos and in which convection ensues due to compositional gradients between the low-entropy PNS core ($r<10$\,km, $s\approx 1\,k_{\rm b}/{\rm b}$, $Y_e\approx0.25$) and the inner disk ($r>35$\,km, $s\approx 7\,k_{\rm b}/{\rm b}$, $Y_e\lesssim0.15$). Magnetic and convective stresses quickly redistribute angular momentum within the first $\lesssim\!50\,{\rm ms}$, rendering the envelope nearly pressure supported, isentropic, and rigidly rotating with the Keplerian angular velocity at 35\,km, $\Omega_{\rm env}\sim\Omega_{\rm kep}(35\,{\rm km})$ (Fig.~\ref{fig:xz}). The MRI thus ceases to be active in the bulk of this inner region and cannot sustain the generation of a large-scale field and Poynting flux on the timescales of interest for the explosion and for GRB jet formation ($t\sim 0.1-10$\,s). Furthermore, shear between the PNS core and the envelope redistributes angular momentum further in on a timescale of $\sim\!100$\,ms, leading to nearly rigid rotation out to $r\approx 35$\,km by $t\lesssim 200$\,ms (Fig.~\ref{fig:xz}). Even though the PNS envelope is convective (in contrast to NS merger remnants \cite{fields2025magnetic, radice_ab-initio_2023}), the strong toroidal field generated in the first tens of milliseconds post-bounce remains buried near the PNS core. However, kinetic energy of the shear flow in the accretion disk remains available, which is continuously replenished as matter is accreted and the disk viscously spreads.

\textit{Large-scale MRI $\alpha\Omega$ dynamo in the accretion disk---}The magnetized accretion disk surrounding the PNS is cooled by neutrinos, reaching a steady state characterized by a scale-height of $H/r\gtrsim0.25$ and $\beta^{-1}\lesssim0.1$ (see Fig.~\ref{fig:xz}). MRI-driven stresses dominate angular momentum transport as soon as the instability is properly resolved (see End Matter). The azimuthally averaged toroidal magnetic field in the disk shows quasi-periodic reversals of global polarity as captured by the ``butterfly'' diagram with multiple dynamo cycles (Fig.~\ref{fig:dynamo}), similar to other MRI dynamos in simulations of stratified, radiatively efficient accretion \citep{hogg_influence_2018,siegel_three-dimensional_2018,siegel_collapsars_2019,combi_jets_2023-2,kiuchi_large-scale_2024}. 
The dynamo quasi-periodicity settles to $\approx\!50-70$\,ms ($\approx\!10$ orbital periods at $60$\,km), consistent with MRI-driven $\alpha\Omega$ dynamos in stratified media \cite{guan_radially_2011,simon_resistivity-driven_2011,flock_large-scale_2011,hogg_influence_2018}.

\begin{figure}
    \centering
    \includegraphics[width=1.0\linewidth]{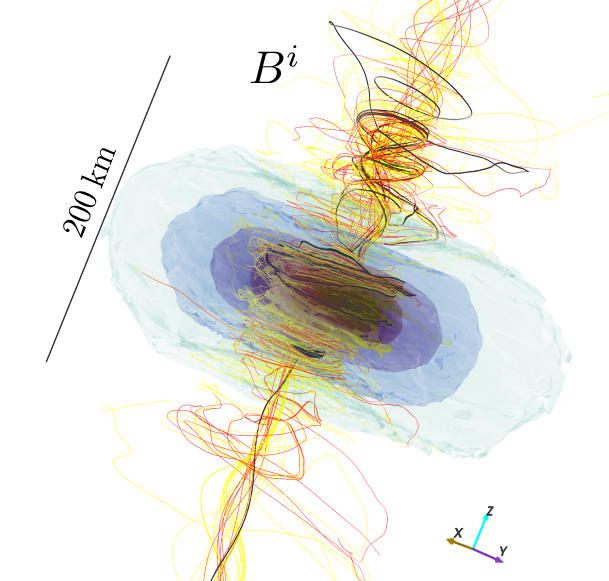}
    \caption{Magnetic field lines seeded at iso-density surfaces of $\rho\approx \lbrace  10^{11}, 10^{12}, 10^{13}\rbrace\,{\rm g}\,{\rm cm}^{-3}$ color-coded as red, yellow, and black, respectively.
    A helical large-scale field develops above the PNS, which anchors to the convective envelope, with some flux tubes connecting directly to the PNS core.}
    \label{fig:3d}
\end{figure}

Toroidal field bundles of coherent polarity buoyantly emerge from the disk at $\approx\!2H$ when $\beta^{-1}\sim 1$, as expected for MRI dynamos \cite{guan_radially_2011,simon_resistivity-driven_2011} (Fig.~\ref{fig:dynamo}). Those bundles close to the PNS envelope migrate latitudinally to the polar cap as they buoyantly rise from the disk, anchor above the PNS, and connect with the convective envelope (Fig.~\ref{fig:dynamo}). When the magnetic pressure increases to $\beta^{-1} \gtrsim 1$, hoop stresses acting at the base of the polar region give rise to a large-scale helical structure with net vertical flux concentrated within a cylindrical region of radius $\approx\!10\,{\rm km}$  (Fig.~\ref{fig:3d}). The `magnetic tower' extracts energy outward and forms a jet-like structure, reminiscent of those arising above the remnant in neutron-star mergers \cite{combi_jets_2023-2}. 


\begin{figure}
    \centering
    \includegraphics[width=1\linewidth]{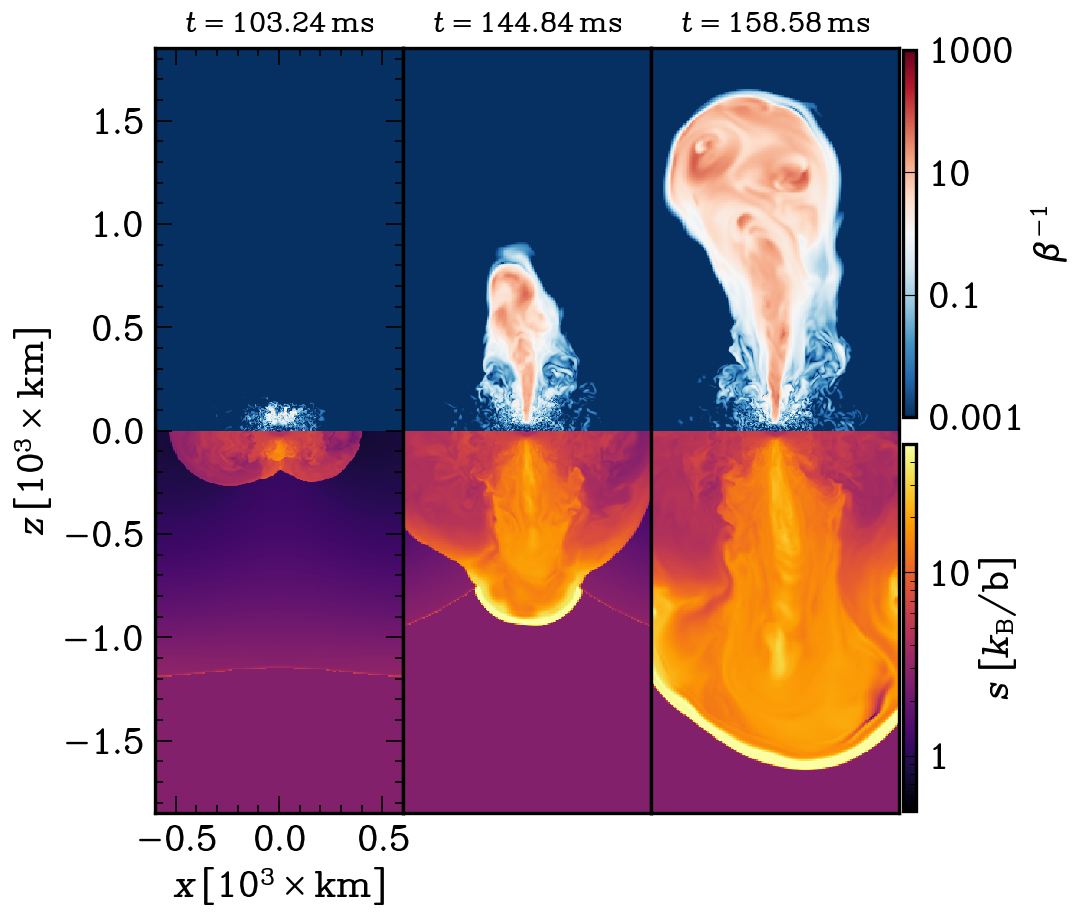}
    \caption{Snapshots of magnetic to fluid pressure ($\beta^{-1}$; top) 
    and entropy per baryon (bottom) in the $xz$-plane. A jet is launched from the PNS at $t\approx 100\,{\rm ms}$ and plows through the collapsing WD until it breaks out from the stellar surface at $t\approx 140\,{\rm ms}$, driving a shock into the circumstellar medium.}
    \label{fig:jet}
\end{figure}

\textit{Breakout, winds, and stripes.---}A collimated, magnetized and baryon-loaded ($\beta^{-1}\gtrsim100$; magnetization $\sigma\sim 0.1$), mildly relativistic ($u^r\sim 0.3\,c$) outflow propagates along the polar axis through the infalling material, generating a high-entropy ($s\sim 10 k_{\rm B}/\text{baryon}$) polar funnel. Baryon-loading arises due to mass-ablation caused by winds from the PNS and disk, but the jet carries a substantial Poynting flux of $\gtrsim\!10^{50}\,{\rm erg}\,{\rm s}^{-1}$ over the timescales probed here. The jet head plows through the remaining envelope of mass $\Delta M=10^{-2}\,{M_{\odot}}$ and extent $\Delta r\sim 1000\,{\rm km}$ fast enough to avoid developing disruptive kink instabilities (Fig.~\ref{fig:jet}). This differs from the magnetorotational collapse of massive stars, in which the extended, massive envelope may hinder the propagation of jets and winds \citep{mosta2014Magnetorotational}. Shock breakout at the stellar surface occurs at around $t\approx 140$\,ms, after which the jet drives a strong shock into the circumstellar medium (Fig.~\ref{fig:jet}), potentially giving rise to observable afterglow emission.


Together with the jet, slower ($u^r\sim0.1\,c$) MHD winds ($\beta^{-1}\sim 1$) are launched from the disk as a result of a heating-cooling imbalance in the corona \cite{siegel_three-dimensional_2018} and by global magnetic stresses. 
Mass ejection of unbound material from the disk starts as soon as the jet breaks out of the star and reaches a steady state of $\dot{M}_{\rm out} \approx 10^{-1}\,M_{\odot}\,{\rm s}^{-1}$ (measured at $r=300\,{\rm km}$) when most of the outer layers of the collapsing star in polar directions have been accreted or blown up ($t\approx 170\,{\rm ms}$). By the end of the simulation, the winds have ejected $M_{\rm ej}\approx10^{-2}\,{M_{\odot}}$ and the remaining disk mass is $M_{\rm disk} \approx 0.15\,{M_{\odot}}$. The mass distribution of the ejecta peaks at $Y_{\rm e}\approx 0.25$, with a tail to $Y_{\rm e} \lesssim 0.2$, which together with the entropy and expansion timescales (see above and End Matter) results in the production of r-process elements primarily below the 2nd abundance peak (atomic mass number $A \lesssim 130$), but also of a sizable amount of heavier lanthanide bearing material with a higher opacity \cite{lippuner_r-process_2015}.

The supply of buoyant toroidal bundles of alternating polarity to the polar region forces reconnection of the vertical magnetic flux, producing a `striped' jet, similar to those generated in other contexts \citep{drenkhahn02,christie_role_2019, kaufman_striped_2023,mahlmann_striped_2020}. During flux advection of the second dynamo cycle at $\approx\! 170\,{\rm ms}$ (Fig.\ref{fig:dynamo}, top, third panel), the jet's magnetization drops to $\beta^{-1}\lesssim$ 0.1 and its Poynting flux decreases by 50\%. The polarity of the vertical flux is not completely inverted, and the jet regains full power in the third dynamo cycle at $\approx\! 210\,{\rm ms}$. The imprint of the dynamo cycle in the periodicity of the Poynting flux self-consistently obtained here is a direct expectation of an MRI-driven $\alpha\Omega$ dynamo. We expect these cycles to continue with increasing period as the disk viscously spreads over time, until the jet becomes dominated by a single polarity over timescales of interest.


\textit{Implications.---}We demonstrate that the magnetized, neutrino-cooled accretion disk formed around the PNS can generate large-scale flux to launch and sustain a magnetized jet. However, to power a GRB, the magnetization of the jet, limited by baryon pollution from the disk/PNS winds throughout the simulation, must increase to $\sigma \gtrsim 100$ to eventually reach relativistic speeds at infinity, $\Gamma_{\infty}\sim \sigma$, consistent with compactness constraints on GRB emission \citep{lithwick_lower_2001}. Provided that a strong, large-scale flux remains anchored to the PNS as it cools, a magnetar forms and such a rise in $\sigma$ is naturally expected as disk material is accreted or ejected, the disk viscously spreads, and the neutrino-heated outflow from the PNS surface polluting the jet abates \citep{thompson_millisecond_2005,metzger_proto-magnetar_2011}.  Given a disk-laden proto-magnetar system similar to those detailed above, Ref.~\citep{metzger_magnetar_2018} predicts that the jet achieves $\sigma \sim 100$ on timescales of $\sim\!10\,{\rm s}$.  The long-term behavior of the PNS field, however, remains an open question. 

In addition to potentially powering a GRB, we demonstrate the self-consistent extraction of a substantial amount of neutron-rich material through magnetized winds, which may give rise to a luminous kilonova and r-process elements, as suggested previously \cite{batziou_nucleosynthesis_2024,cheong_gamma-ray_2025}. Although our disk outflows are neutron-rich, as the accretion disk continues to viscously spread in time, eventually neutrino irradiation from the PNS may raise a greater fraction of the disk outflows to $Y_e \gtrsim 0.5$ sufficient to synthesize $^{56}$Ni in addition to $r$-process nuclei \cite{Metzger+09AIC}.  Because of the larger specific radioactive heating rate of $^{56}$Ni than $r$-process nuclei \cite{siegel_collapsars_2019}, even a modest $^{56}$Ni mass fraction could generate a distinctive light-curve shape that would distinguish its kilonova from a neutron-star merger.

One possible point of tension in invoking AIC as the source of recent long GRBs with kilonovae \citep{rastinejad_kilonova_2022-1,Levan+24} comes from overall energetics. The rotational energy in the system at the end of our simulation is still large, $E_{\rm rot}\approx 0.9\times10^{52}\,{\rm erg}$, and will likely grow further as the PNS contracts as it cools to its final state. Most of this energy must be eventually released into the environment, whether contributing to the kinetic energy of the disk outflows or an ultra-relativistic jet.  The synchrotron radio emission produced by the collision of this material with the interstellar medium for $E_{\rm rot}\sim 10^{52}\,{\rm erg}$ should be substantially brighter than other merger-powered GRBs \citep{metzger_constraints_2014-2,fong_radio_2016,schroeder_late-time_2020}. 

The disk-driven turbulent and mean-field dynamos identified here may act in a similar form in other systems with accreting PNSs, such as in magnetorotational core-collapse, where previous envelope-based dynamos \cite{mosta_large-scale_2015} may be insufficient to explode the star and sustain a jet over timescales of interest (see above), and binary neutron-star mergers, where recent results for jet formation indicate similar dynamo properties \cite{combi_jets_2023-2,kiuchi_large-scale_2024}.


\acknowledgments

We thank Chris Thompson, Alexander Philippov, Michael M\"uller, and Philipp M\"osta for fruitful discussions. This research was enabled in part by support provided by SciNet (www.scinethpc.ca) and Compute Canada (www.computecanada.ca). The authors gratefully acknowledge the computing time made available to them on the high-performance computer ``Lise'' at the NHR Center NHR@ZIB. This center is jointly supported by the German Federal Ministry of Education and Research and the state governments participating in the NHR (www.nhr-verein.de/unsere-partner). L.C.~is a CITA National fellow and  acknowledges the support by the Natural Sciences and Engineering Research Council of Canada (NSERC), funding reference DIS-2022-568580. D.M.S.~acknowledges the support of NSERC, funding reference number RGPIN-2019-04684. D.M.S.~acknowledges a Visiting Fellow position at Perimeter Institute. This research was supported in part by Perimeter Institute for Theoretical Physics. Research at Perimeter Institute is supported in part by the Government of Canada through the Department of Innovation, Science and Economic Development Canada and by the Province of Ontario through the Ministry of Colleges and Universities. B.D.M. acknowledges support from the National Science Foundation (grant AST-2406637) and the Simons Foundation (grant 727700).  The Flatiron Institute is supported by the Simons Foundation.

\bibliographystyle{apsrev4-2}
\bibliography{ms,astro,gr-astro}

\appendix

\section{End Matter}

\subsection{Simulation set-up}

The current version of our GRMHD code has several modifications with respect to previous versions \cite{mosta_grhydro_2014, siegel_three-dimensional_2018,combi_grmhd_2023}. Spacetime is evolved with \texttt{CTGamma} \cite{pollney_high_2011}, which implements the Z4c formulation \cite{bernuzzi_constraint_2010}. The magnetic field is evolved with a newly implemented vector potential formalism in the generalized Lorenz gauge \cite{farris_binary_2012} using an upwind constrained transport scheme \citep{del_zanna_echo_2007}. We use WENO-Z reconstruction and an HLLE Riemann solver for the hydrodynamic variables. Recovery of primitive variables from conservative ones is handled by the framework of Ref.~\cite{siegel_recovery_2018}. We use the finite-temperature, composition-dependent LS220 equation of state \cite{lattimer_generalized_1991} in tabulated form \cite{schneider_open-source_2017}. A density floor is set at $10^2\,{\rm g}\,{\rm cm}^{-3}$. We use the Carpet driver \cite{schnetter_evolutions_2004} for Berger-Oliger mesh refinement with subcycling in time. 
As the star collapses, we successively activate new refinement levels of our grid hierarchy until nuclear densities are reached. 

\subsection{Initial conditions for the magnetic field and caveats}

\begin{figure}[tb]
    \centering
    \includegraphics[width=1.0\linewidth]{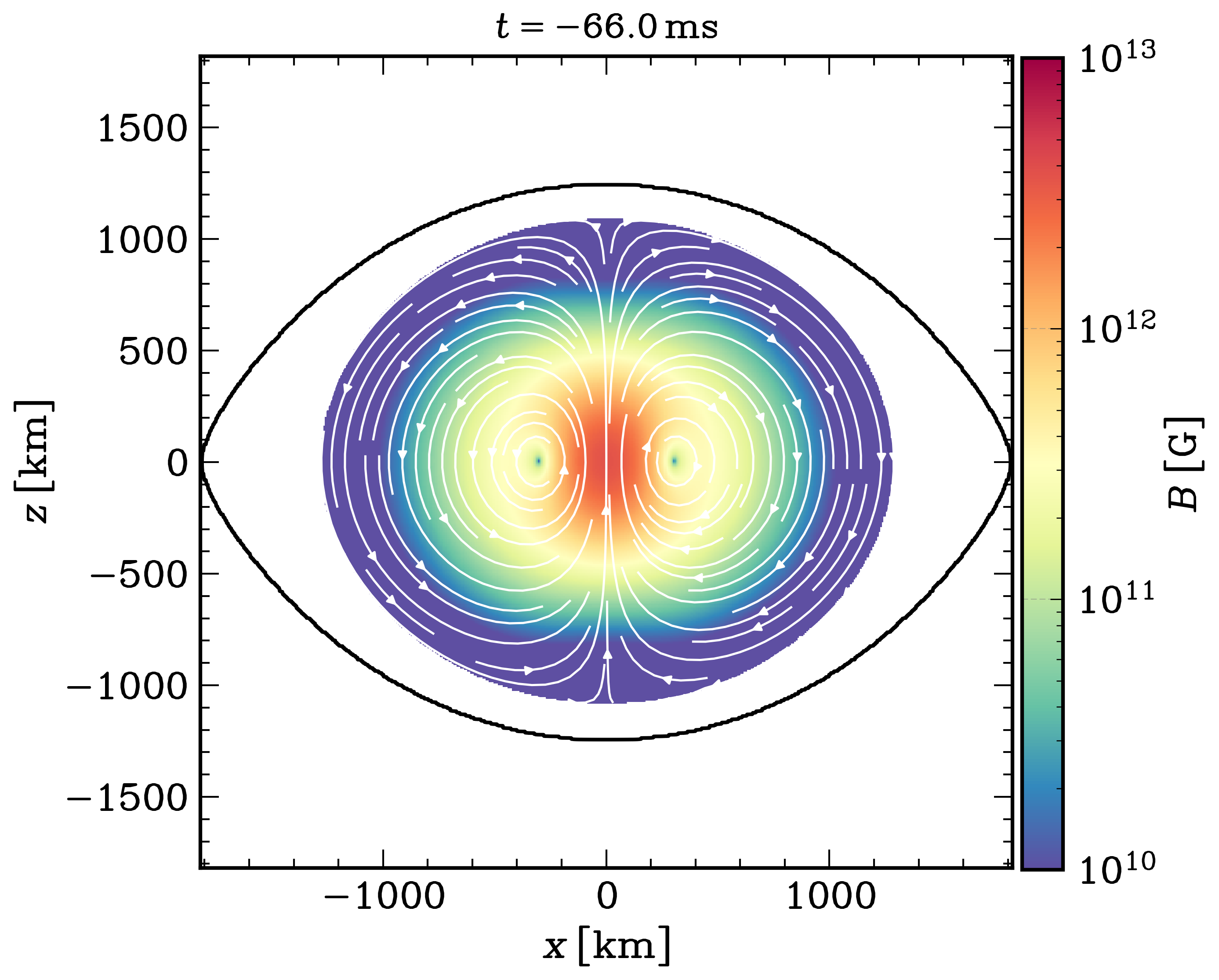}
    \includegraphics[width=1.0\linewidth]{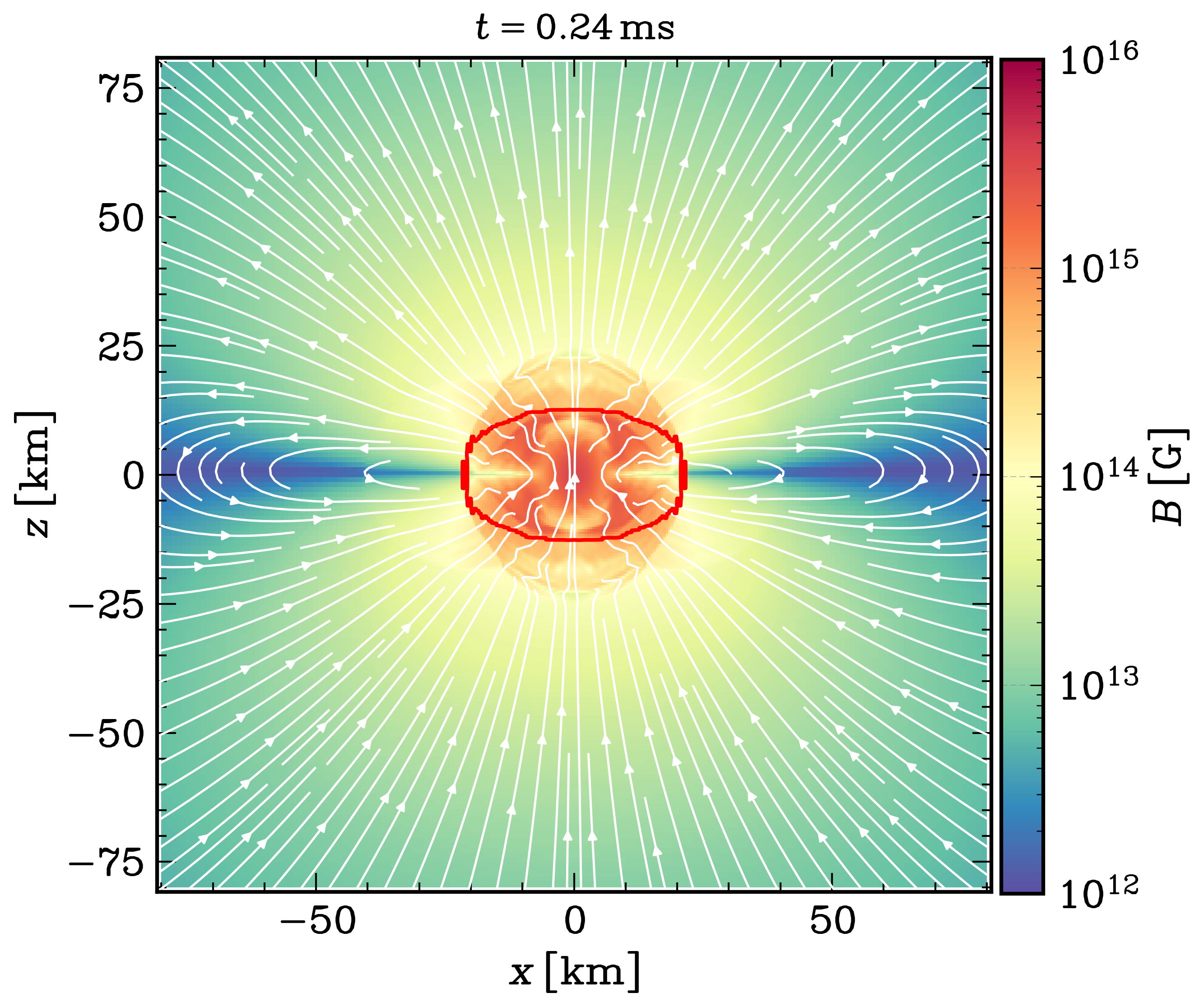}
    \caption{Top: Initial conditions for the magnetic field strength in the $xz$-plane at the beginning of the simulation. Bottom: magnetic field strength in the $xz$-plane at bounce. White lines represent magnetic field lines, the black contours represent the surface of the star, and the red contours mark the isodensity surface at $10^{12}\,{\rm g}\,{\rm cm}^{-3}$.}
    \label{figapp:1}
\end{figure}

We start the simulation with a strong but highly compact poloidal seed magnetic field in the interior of the collapsing WD, as shown in Fig.~\ref{figapp:1}. This setup is (a) chosen to avoid significant influence of the topology on jet production and (b) motivated by the fact that WDs with large-scale fields (i.e., those with a strong magnetosphere) are unlikely to accrete substantial angular momentum and will not be rapidly rotating at the time of collapse \citep{thompson95}, as we assume here. This setup differs from previous 2D axisymmetric AIC and 3D iron-core collapse simulations \cite{dessart_magnetically_2007, cheong_gamma-ray_2025,bugli_three-dimensional_2021,powell_three_2023}, in which a vertical field is set nearly constant up to radii of $\sim1000\,{\rm km}$, resulting in PNSs with exterior large-scale, magnetar-strength fields by flux compression and typically two orders of magnitude larger $E_{\rm mag}$ for the same initial maximum field strength.  As we show below, our initial conditions give rise to an initially dynamically irrelevant field, which is subsequently self-consistently amplified by turbulent and mean-field dynamos driven by the MRI. 

\begin{figure}[tb]
    \centering
    \includegraphics[width=1.0\linewidth]{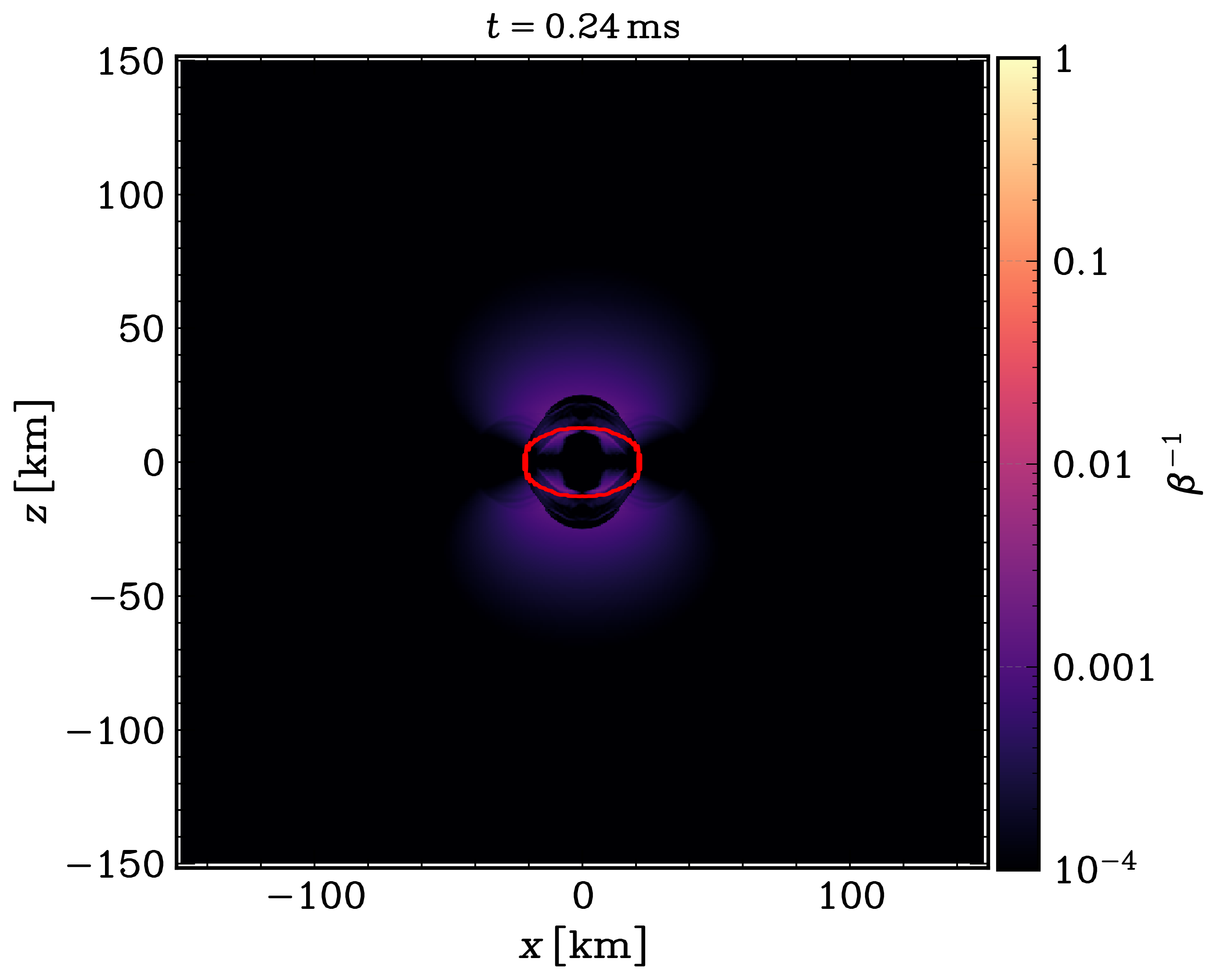}
    \includegraphics[width=1.0\linewidth]{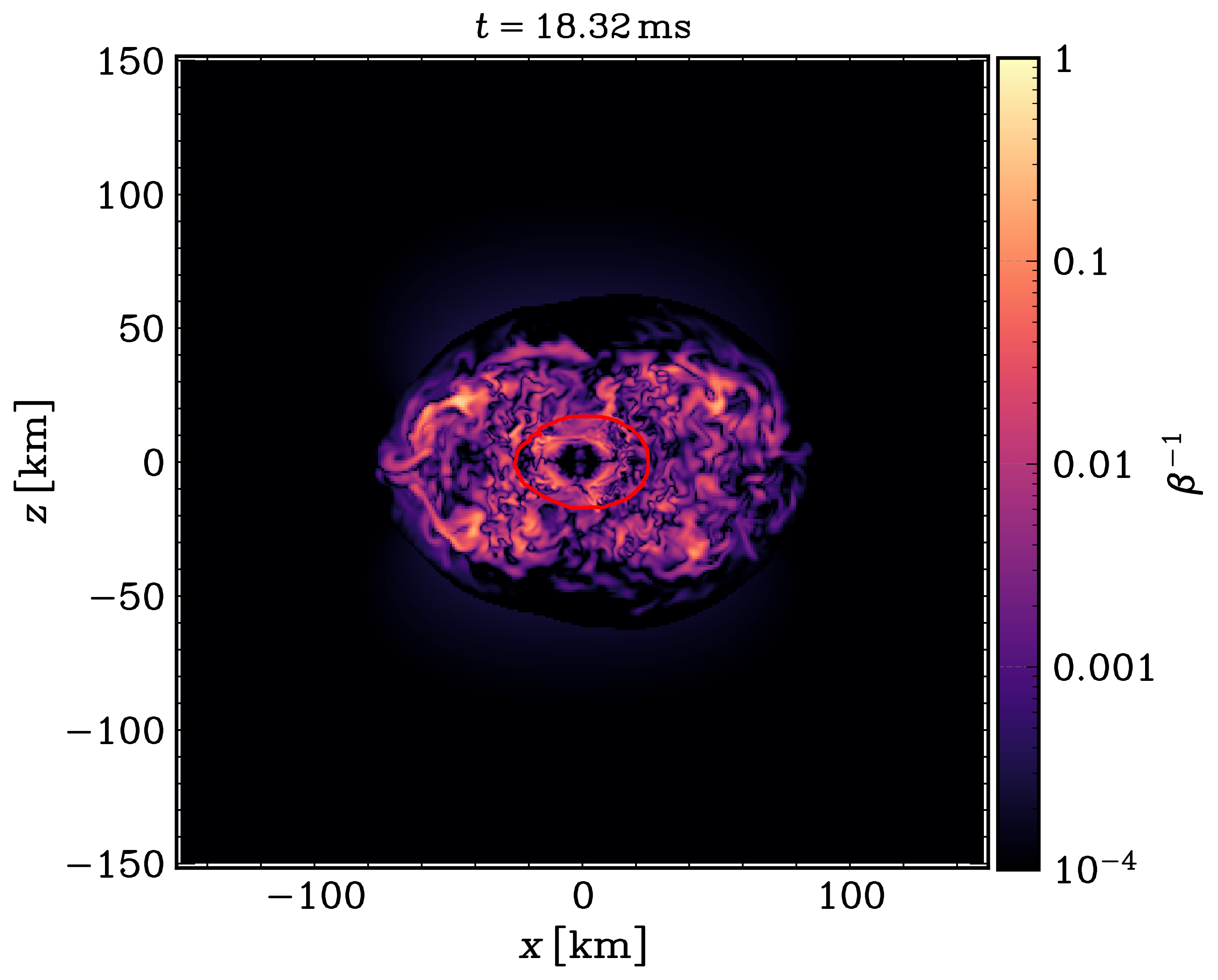}
    \caption{Snapshots of the inverse plasma-$\beta$ parameter, $\beta^{-1}:= b^2/2p$, in the $xz$-plane at bounce (top), when the field is weak and dynamically irrelevant, and at $\approx 18.32\,{\rm ms}$ post bounce (bottom).}
    \label{figapp:2}
\end{figure}

Because the initial field is not homogeneous in space, flux-freezing amplification occurs mainly near the center, reaching $B\approx 10^{15}$\,G at bounce. Because the collapse is not isotropic owing to rotation, the initial dipole field preferentially compresses along the vertical $z$-axis, forcing partial reconnection in the equatorial plane, where the field strength drops as a result (see Fig.~\ref{figapp:1}). At bounce, the field in the inner shear layer outside the PNS core (what develops into the envelope) is $B\approx10^{14}\,{\rm G}$ at maximum and weak compared to the fluid pressure ($\beta^{-1} \lesssim10^{-3}$), see Figs.~\ref{figapp:1} and \ref{figapp:2}.  After bounce ($t\approx 20\,{\rm ms}$), the MRI instability operates in the shear flow around the PNS core, driving the magnetic fields to $\beta^{-1} \gtrsim 0.1$ (Fig.~\ref{figapp:2}). The resulting field is predominantly turbulent, with a net toroidal flux in the inner shear flow surrounding the PNS core. 

Previous two-dimensional (axisymmetric) MHD simulations of AICs \citep{dessart_magnetically_2007, cheong_gamma-ray_2025} are limited by the anti-dynamo theorem in axisymmetry \cite{cowling_magnetic_1933}. These use an initial magnetic seed field with similar maximum field strength, but reach dynamically relevant fields ($\beta^{-1}\approx 1$) outside the PNS just after bounce purely by flux compression due to the different size of the initial loop.

Our study is mainly limited by the finite resolution of global simulations, which sets a minimum initial field strength in the shear layer around the PNS to resolve the MRI soon after bounce (see also below). However, since the MRI is a weak-field instability leading to exponential amplification and given the enormous free energy in differential rotation, even much weaker (and more realistic) initial magnetic seed fields in the collapsing WD than the one employed here will be sufficiently amplified by the MRI to yield similar results than those presented here. Assuming that a field strength of $B_{\rm disk}\approx 10^{15}$\,G needs to be generated by the MRI in a disk region of rotation period $T = 2\pi / \Omega \sim 1-2\,\text{ms}$ (Fig.~\ref{fig:ek_evolmag}) within $t_{\rm amp}\approx 50$\,ms (Fig.~\ref{fig:dynamo}) to generate a jet on similar timescales as shown here, as well as a flux amplification factor of $f_A \approx 100$ during collapse, a maximum pre-collapse field strength within the WD core of
\begin{equation}
    B_0 \approx 30\,\text{mG}\left(\frac{B_{\rm disk}}{10^{15}\text{G}}\right) \left(\frac{f_A}{100}\right)^{-1}\mskip-10mu\exp\left[\frac{t_{\rm amp}/50\text{ms}}{T/1.5\text{ms}}\right]^{-1}
\end{equation}
is sufficient---orders of magnitude below observational constraints on WD exterior field strengths ($\lesssim 10^9$\,G) \cite{ferrario_magnetic_2005} as well as below maximum field strengths ($\lesssim 10^6$\,G) reached by dynamos during the crystallization of the WD core \cite{schreiber_origin_2021,ginzburg2022slow}.

The saturation strength of the magnetic field and the small-scale dynamo likely depend only weakly on the initial seed field, as long as diffusive/resistive effects are small. This is indeed expected for the accretion disk/shear flow, whereas the effect of neutrino viscosity deep inside the PNS (strongly diffusive regime) can be substantial for field strengths smaller than $B\gtrsim10^{11}\,{\rm G}$ \cite{guilet_neutrino_2015}. 
The saturation strength through parasitic modes only weakly depends on the magnetic Prandtl number \cite{pessah_saturation_2009,Reboul-Salze:2021rmf}, which is controlled here by the grid resolution ($P_{\rm m}\sim 1$). The magnetic Prandtl number may also influence the $\alpha\Omega$ dynamo in that the magnetic stresses are enhanced \cite{simon_resistivity-driven_2011}, leading to shorter dynamo periods and faster viscous spreading of the disk. However, properties of the $\alpha\Omega$ dynamo in the $P_{\rm m}\gtrsim 1-100$ regime are moderate and scale weakly with $P_{\rm m}$, and start to saturate around $P_{\rm m}\approx 60$ \cite{simon_resistivity-driven_2011,held_mri_2024,reboul2025mri}---qualitative changes are not expected.

Although the magnetic field here amplifies self-consistently due to the MRI at all scales and stages, it might still carry imprints of the initial large-scale loop; we leave a study of the impact of the initial magnetic field topology and strength in the early post-bounce phase to future work. 



\begin{figure}[tb]
    \centering
    \includegraphics[width=1\linewidth]{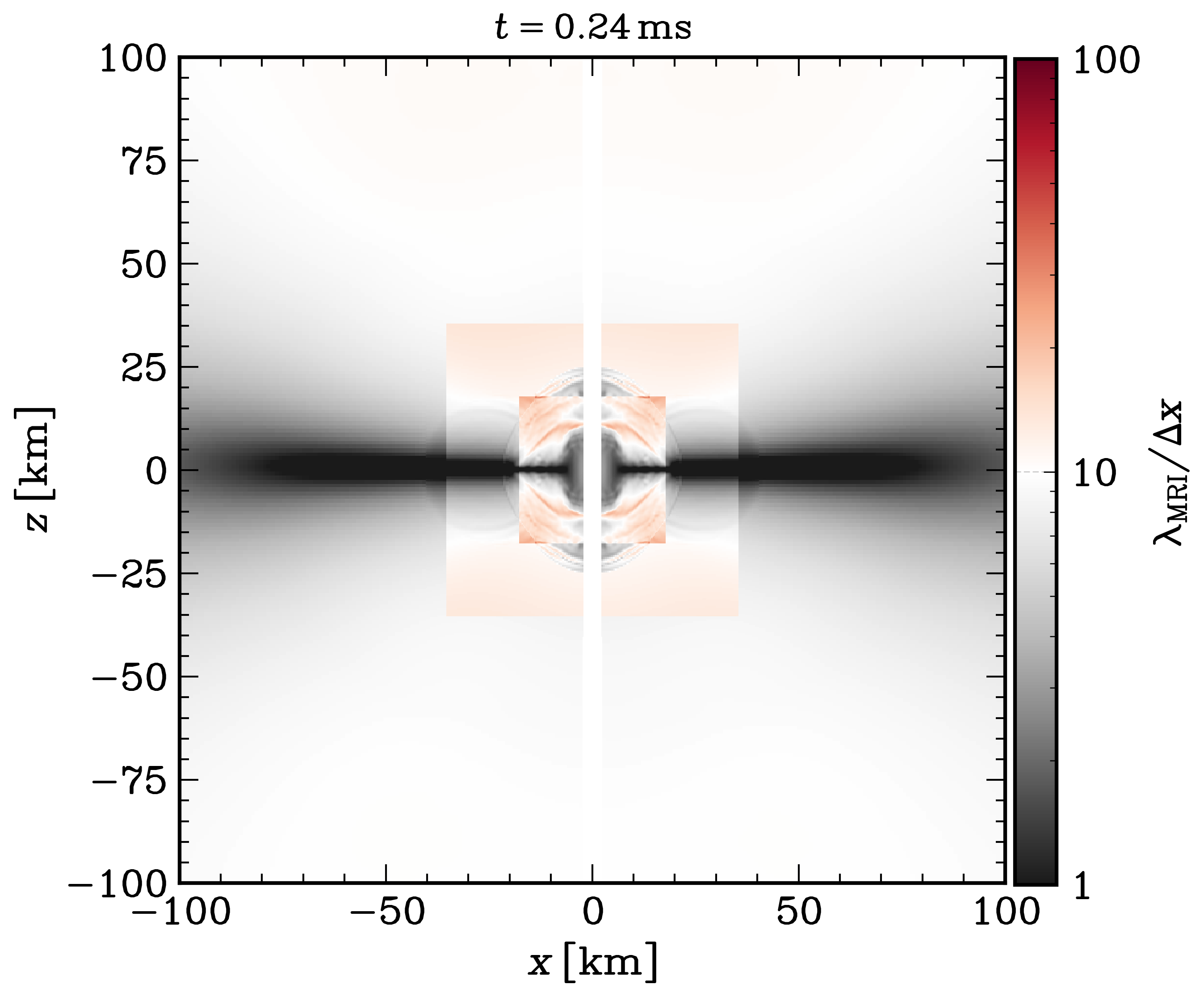}
    \includegraphics[width=1\linewidth]{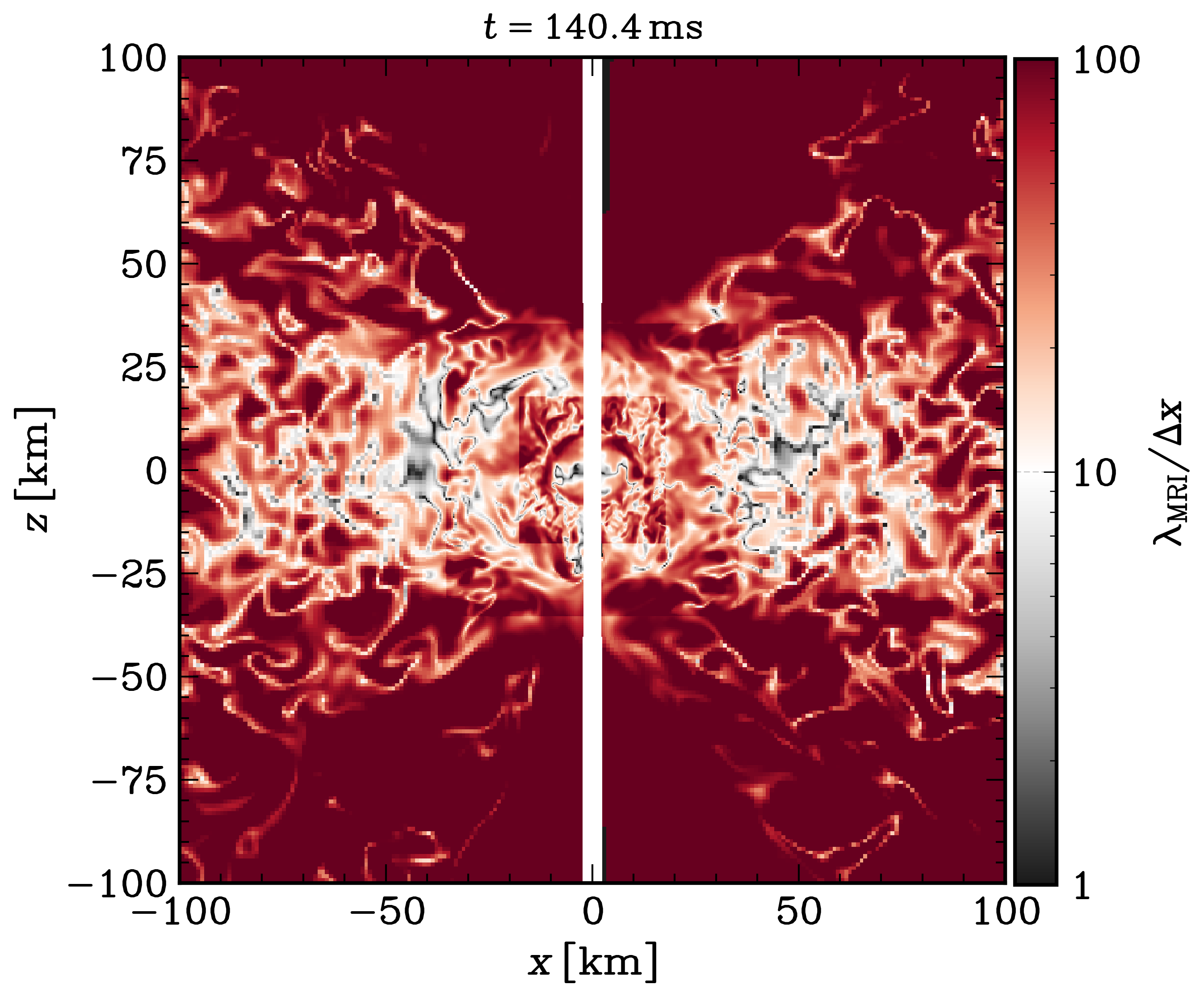}
    \caption{MRI quality factor just after the bounce (top) and once the accretion disk has been fully established (bottom).}
    \label{fig:lambda}
\end{figure}

\begin{figure*}[htb!]
    \centering
    \includegraphics[width=1.0\linewidth]{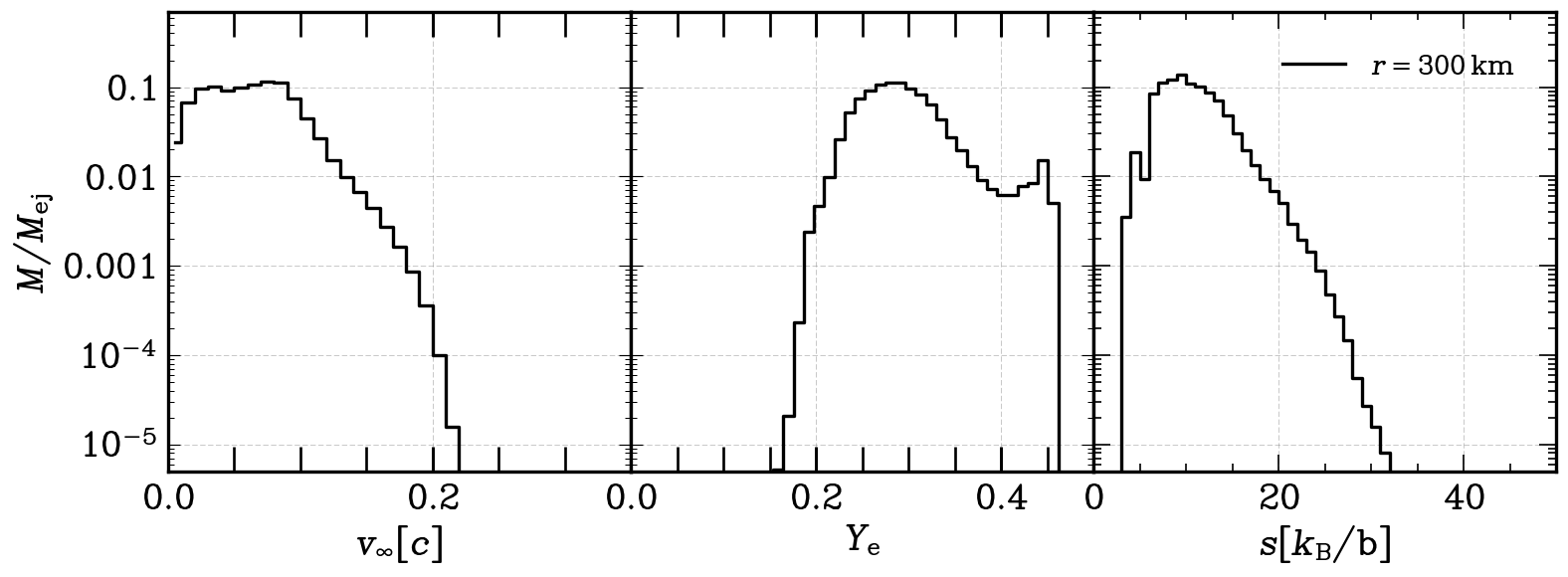}
    \caption{Mass distribution of the cumulative unbound, ejected material, measured at $r\approx300\,{\rm km}$ as a function of asymptotic velocity $v_{\infty}$, electron fraction $Y_{\rm e}$, and entropy $s$.}
    \label{fig:histogram}
\end{figure*}

\subsection{Resolving the magnetorotational instability}

The quality to resolve the fastest growing mode of the MRI can be monitored by computing $\lambda_{\rm MRI}/\Delta x$, where $\Delta x$ is the resolution of the grid at a given location and 
\begin{equation}
    \lambda_{\rm MRI} = \frac{\Omega}{2\pi} \frac{b}{\sqrt{\rho h+b^2}}
\end{equation}
is the wavelength of the fastest growing MRI mode, with $h$ being the specific enthalpy. Typically, $\lambda_{\rm MRI}/\Delta x\gtrsim 10$ is required to properly resolve the instability \cite{siegel_magnetorotational_2013}.

At bounce, the instability is resolved near the shear layer with more than $10$ cells, except near the equatorial plane, where the field strength is much smaller due to the collapse dynamics (see above; Fig.~\ref{fig:lambda}, top panel). However, as the fluid becomes turbulent and mixes, the instability is well resolved throughout the simulation domain where it is active and at all times (Fig.~\ref{fig:lambda}, bottom).

\subsection{Properties of the outflows}

The system ejects a considerable amount of mass in PNS neutrino-driven winds, MHD-thermal disk winds, and the magnetized jet. The total mass ejected in unbound $-hu_t>h_{\infty}$ outflows, where $h_{\infty}$ is the specific enthalpy at infinity and $u_t$ the covariant time component of the four-velocity, as measured at a sphere of $r\approx300\,{\rm km}$ is $M_{\rm ej}\approx10^{-2}\,{M_{\odot}}$ by the end of evolution at $t\approx250\,{\rm ms}$. 

The properties of the unbound outflow are represented in the mass distribution histograms of Fig.~\ref{fig:histogram}. Most of the mass is uniformly ejected at small velocities with a peak at $v\approx0.1\,c$ and a tail up to $v\gtrsim 0.2\,c$, as expected for winds from neutrino cooled accretion disks \cite{siegel_three-dimensional_2018,fernandez2019Longterm}. The composition of the ejecta is relatively neutron-rich, peaking at $Y_{\rm  e}\approx0.25$, with a tail to $Y_{\rm e}\lesssim 0.2$. The entropy peaks at $s\approx10\,k_{\rm b}/{\rm b}$, with a tail up to $s\gtrsim 30 \,k_{\rm b}/{\rm b}$.



\end{document}